\documentclass[a4paper]{jpconf}
\usepackage{graphicx}
\begin{document}
\title{An anti-Schwarzshild solution:\\
wormholes and scalar-tensor solutions}

\author{Jos\'e P. Mimoso and Francisco S. N. Lobo}

\address{Centro de Astronomia e Astof\'{\i}sica da Universidade de Lisboa, Avenida Professor Gama
Pinto 2, P-1649-003 Lisbon, Portugal}

\ead{jpmimoso@cii.fc.ul.pt, flobo@cii.fc.ul.pt}

\begin{abstract}
We investigate a static solution with an hyperbolic nature,
characterised by a pseudo-spherical foliation of space. This
space-time metric can be perceived as an anti-Schwarzschild
solution, and exhibits repulsive features. It belongs to the class
of static vacuum solutions termed ``a degenerate static solution
of class A'' (see [1]). In the present work we review its
fundamental features, discuss the existence of generalised
wormholes, and derive its extension to scalar-tensor gravity
theories in general.

\end{abstract}

\section{Introduction}

We consider a largely ignored metric which belongs to a class of
vacuum solutions referred to as degenerate solutions of class A
\cite{Stephani:2003tm} given by
\begin{equation}
ds^2= -e^{\mu(r)}\,{\rm d}t^2+ e^{\lambda(r)} \,{\rm
d}r^2+r^2\,({\rm d}u^2+\sinh^2{u}\,{\rm d}v^2) \,,
\label{metric_constnc}
\end{equation}
which are axisymmetric solutions \cite{Bonnor & Martins 1991}, but
where the usual 2-dimensional spheres are replaced by
pseudo-spheres, ${\rm d}\sigma^2={\rm d}u^2+\sinh^2{u}\,{\rm
d}v^2$, i.e., by surfaces of negative, constant curvature. These
are still surfaces of revolution around an axis, and $v$
represents the corresponding rotation angle. For the vacuum case
we get
\begin{equation}
e^{\mu(r)}=e^{-\lambda(r)} =\left(\frac{2\mu}{r}-1\right) \; ,
\label{metric_constnc-antiScwarz}
\end{equation}
where $\mu$ is a constant \cite{Stephani:2003tm,Bonnor & Martins
1991}. We immediately see that the static solution holds for
$r<2\mu$ and that there is a coordinate singularity at $r=2\mu$
(note that $|g|$ neither vanishes nor becomes $\infty$ at
$r=2\mu$)\cite{Anchordoqui:1995wa}. This is the complementary
domain of the exterior Schwarzschild solution. In the region
$r>2\mu$, as in the latter solution, the $g_{tt}$ and $g_{rr}$
metric coefficients swap signs. Defining $\tau=r$ and $\rho=t$, we
obtain
%\begin{equation}
${\rm d} s^2= -{\rm d}\tilde\tau^2+A^2(\tilde{\tau})\,{\rm
d}\rho^2 +B^2(\tilde\tau)\,({\rm d}u^2+\sinh^2{u}\,{\rm d}v^2),$
%\end{equation}
with the following parametric definitions $\tilde{\tau} = -\tau
+2\mu \, \ln|\tau-2\mu|$, $A^2=2\mu/\tau-1$ and
$B^2(\tau)=\tau^2$, which is a particular case of a Bianchi III
axisymetric universe.

Using pseudo-spherical coordinates $\{x = r\sinh u  \cos v,\, y =
r\sinh u  \sin v,\,z = r \cosh u,\,w = b(r)\}$, the spatial part
of the metric (\ref{metric_constnc}) can be related to the
hyperboloid $w^2+x^2+y^2-z^2 = \left(b^2/r^2 - 1\right)\, r^2$
embedded in a 4-dimensional flat space. We then have
\begin{equation}
{\rm d}w^2 +{\rm d}x^2+{\rm d}y^2-{\rm d}z^2 = \left[(b'(r))^2 -
1\right]\,{\rm d}r^2+ r^2 \left( {\rm d}u^2+\sinh^2u\, {\rm d}v^2
\right) \; .
\end{equation}
where the prime stands for differentiation with respect to $r$,
and $b(r) = \mp 2\sqrt{2\mu}\sqrt{2\mu-r}$. We can recast metric
(\ref{metric_constnc}) into the following
\begin{eqnarray}
{\rm d} s^2= -\, \tan^2\left[\ln\left( \bar r\right)^{\mp 1}
\right] \,{\rm d}\tau^2+
 \left(\frac{2\mu}{\bar r}\right)^2\,
 \cos ^4\left[\ln\left( \bar r\right)^{\mp 1}\right]\, \left[{\rm d}\bar{r}^2\,+ \bar r^2\,({\rm d}u^2
+\sinh^2{u}\,{\rm d}v^2) \right]\; , \label{Isotrop3}
\end{eqnarray}
which is the analogue of the isotropic form of the Schwarzschild
solution. The spatial surfaces are conformally flat, but the flat
metric is not euclidean. Indeed, the 3--dim spatial metric $ {\rm
d}\sigma^2 = {\rm d}\bar{r}^2\,+ r^2\,({\rm d}u^2+\sinh^2{u}\,{\rm
d}v^2) \label{hyperbolic flatness}$ is foliated by 2-dimensional
surfaces of negative curvature, since ${R^2}_{323} = - \sinh^2 u<
0$, and it corresponds to $ {\rm d}\sigma^2 = {\rm d}x^2+ {\rm
d}y^2-{\rm d}z^2$. We thus cannot recover the usual Newtonian
weak-field limit.

Analysing the ``radial'' motion of test particles, we have the
following equation
%\begin{equation}
$\dot{r}^2+ \left(\frac{2\mu}{r} -1\right)\,\left(1+\frac{h^2} {r^2
\sinh^2 u_\ast } \right) = \epsilon
$
%\end{equation}
where $\epsilon$ and $h$ are constants of motion defined by
$\epsilon= \left(2\mu/r -1\right)\, \dot{t} = {\rm const_t}$ and
$h^2= r^2 \,\sinh^2 u_\ast \, \dot{v} = {\rm const_v}$, for fixed
$u=u_\ast$. The former and latter constants represent the energy
and angular momentum per unit mass, respectively. We thus define
the potential
%\begin{equation}
$2V(r) = \left(\frac{2\mu}{r} -1\right)\,\left(1+\frac{h^2} {r^2
\sinh^2 u_\ast } \right).\label{geod_potential}$
%\end{equation}
This potential is manifestly repulsive, crosses the $r$-axis at
$r=2\mu$, and for sufficiently high values of $h$ it has a minimum at
%\begin{equation}
$r_{\pm} =(h^2\mp \sqrt{h^4 - 12 \mu^2 h^2})/(2\mu)$.
% \; ,\end{equation}
%provided the angular momentum per unit mass $h$ takes a high
%enough value.
However this minimum %when it exists
falls outside the $r=2\mu$ divide. So a test particle is subject
to a repulsive potential
%and its radial coordinate is ever increasing
forcing it to inevitably cross the event horizon at $r=2\mu$
attracted either by some mass at the minimum or by masses at
infinity. In \cite{Bonnor & Martins 1991} it is hinted that the
non-existence of a clear Newtonian analogue is related to the
existence of mass sources at $\infty$, but no definite conclusions
were drawn.

\section{Alter-ego of Morris-Thorne wormholes}

A natural extension of the solution
(\ref{metric_constnc-antiScwarz}) would be to add exotic matter to
obtain static and pseudo-spherically symmetric traversable
wormhole solutions \cite{Lobo:2009du}. Consider the metric
(\ref{metric_constnc}) given by $\mu(r)=2\Phi(r)$ and
$\lambda(r)=-\ln[1-b(r)/r]$. The coordinate $r$ decreases from a
constant value $\mu$ to a minimum value $r_0$, representing the
location of the throat of the wormhole, where $b(r_0)=r_0$, and
then it increases from $r_0$ back to the value $\mu$. The
condition $(b/r-1) \geq 0$ imposes that $b(r)\geq r$, contrary to
the Morris-Thorne counterpart \cite{Morris:1988cz}.

The solution provides the following stress-energy scenario
\begin{eqnarray}
\rho(r)&=&-\frac{1}{8\pi} \, \frac{b'}{r^2} \,,
%\label{rhoWH}\,,\\
   \qquad
p_r(r)=\frac{1}{8\pi} \, \left[ \frac{b}{r^3}+2
\left(\frac{b}{r}-1\right) \frac{\Phi'}{r}
\right] \label{radialpWH}\,,\\
p_t(r)&=&\frac{1}{8\pi} \left(\frac{b}{r}-1\right)\Bigg[\Phi ''+
(\Phi')^2 + \frac{b'r+b-r}{2r(b-r)}\Phi'
 +\frac{b'r-b}{2r^2(b-r)} \Bigg] \label{lateralpWH}\,,
\end{eqnarray}
in which $\rho(r)$ is the energy density, $p_r(r)$ is the radial
pressure, $p_t(r)$ is the pressure measured in the tangential
directions.
Note that the radial pressure is always positive at the throat,
i.e, $p_r=1/(8\pi r_0^2)$, contrary to the Morris-Thorne wormhole,
where a radial tension at the throat is needed to sustain the
wormhole. In addition to this, the mathematics of embedding
imposes that $b'(r_0)>1$ at the throat, which implies a negative
energy density at the throat (see \cite{Lobo:2009du} for more
details). This condition is another significant difference to the
Morris-Thorne wormhole, where the existence of negative energy
densities at the throat is not a necessary condition. Several
interesting equations of state were considered in
\cite{Lobo:2009du}, and we refer the reader to this work for more
details.

\section{Pseudo-spherical scalar-tensor solution}

A theorem by Buchdahl \cite{Buchdahl 1959} establishes the
reciprocity  between any static solution of Einstein's vacuum
field equations and a one-parameter family of solutions of
Einstein's equations with a (massless) scalar field. In the
conformally transformed Einstein frame, note that scalar-tensor
gravity theories are described by
%\begin{equation}
$\tilde{S} = \int\, \sqrt{-\tilde g}\,\left[\left[\tilde R -\frac{1}{2}\,
(\nabla\varphi)^2 \right] + \left( 16\pi
G_N/\Phi^{-2}(\varphi)\right)\, L_{Matter}\right]
\label{E_ST_action}
$.
%\end{equation}
In this representation we have GR plus a massles scalar field
which is now coupled to the matter fields. Different scalar-tensor
theories correspond to different couplings. In the absence of
matter we can use Buchdahl's theorem. So, given the metric
(\ref{metric_constnc}), we derive the corresponding scalar-tensor
solution
\begin{eqnarray}
{\rm d} s^2 &=& -\left(\frac{2\mu}{r}-1\right)^B\,{\rm d}t^2+
\left(\frac{2\mu}{r}-1\right)^{-B}\;{\rm d}r^2
  +\left(\frac{2\mu}{r}-1\right)^{1-B}\,r^2\,({\rm d}u^2
+\sinh^2{u}\,{\rm d}v^2) \, , \label{ST_sol_constnc} \\
\varphi(r) &=& \sqrt{\frac{C^2(2\omega+3)}{16\pi}\varphi_0}\,
\ln\left(\frac{2\mu}{r}-1\right) \,,
\end{eqnarray}
%and
%\begin{equation}
%\varphi(r) = \sqrt{\frac{C^2(2\omega+3)}{16\pi}\varphi_0}\,
%\ln\left(\frac{2\mu}{r}-1\right)
%\end{equation}
where $ C^2=(1-B^2)/(2\omega+3)$ and $-1 \le B \le 1$. This
clearly reduces to our anti-Schwarzschild metric
(\ref{metric_constnc}) in the GR limit when $B=1$, and hence $C=0$
implying that $G=\Phi^{-1}$ is constant. Reverting
$\varphi=\int\sqrt{\Phi_0(2\omega+3)/(16\pi)}\,{\rm
d}\ln(\Phi/\Phi_0)$, and the conformal transformation, $g_{ab}=(2\mu/r-1)^{-C}\,\tilde{g}_{ab}$, we can
recast this solution in the original frame in which the
scalar-field is coupled to the geometry and the content is vacuum,
the so-called Jordan frame.

The $r=2\mu$ limit is no longer just a coordinate singularity, but
rather a true singularity as it can be verified from the analysis
of the curvature invariants. In the Einstein frame this occurs
because the energy density of the scalar field diverges likewise
in the Schwarzschild case\cite{David Wands 1993}. 
%The geodesic motion is repulsive in the $r$ coordinate if $0\le B\le 1$, and has an 
%attractor between $r=0$ and $r=2\mu$ when $-1\le B<0$, corresponding to an orbit with 
%constant-$r$. 
Of paramount importance is that, once again, the ST-solution has no Newtonian limit (as its
GR limit does not have one). This implies that the usual
Parametrized Post-Newtonian formalism that assesses the departures
of modified gravity theories from GR does not hold for this class
of metrics (see \cite{MimLobo1}).

\section{Discussion}

We have outlined the exotic features of the vacuum static solution
with a pseudo-spherical foliation of space. We have revealed the
existence of generalised wormholes, and derived its extension to
scalar-tensor gravity theories. A fundamental feature of these
solutions is the absence of a Newtonian weak field limit, which
reminds us of a quotation from John Barrow ~\cite{Barrow:1991fj}
\begin{quote}
{\em The miracle of general relativity is that a purely
mathematical assembly of second-rank tensors should have anything
to do with Newtonian gravity in any limit}.

\end{quote}
\ack

The authors are grateful to Raul Vera and Guillermo A. Gonz\'alez
for helpful discussions. JPM also acknowledges the LOC members, Ruth, Raul, Jesus and Jos\'e for a very enjoyable conference.

\section*{References}
%%%%%%%%%%%%%%%%%%%%%%%%%%%%

\end{document}